# Exploring the Change in Scientific Readability Following the Release of ChatGPT

Abdulkareem Alsudais

College of Computer Engineering and Sciences, Prince Sattam bin Abdulaziz University, Al-Kharj
11942, Saudi Arabia
E-mail: A.Alsudais@psau.edu.sa

**Abstract**

The rise and growing popularity of accessible large language models have raised questions about their impact on various aspects of life, including how scientists write and publish their research. The primary objective of this paper is to analyze a dataset consisting of all abstracts posted on arXiv.org between 2010 and June 7th, 2024, to assess the evolution of their readability and determine whether significant shifts occurred following the release of ChatGPT in November 2022. Four standard readability formulas are used to calculate individual readability scores for each paper, classifying their level of readability. These scores are then aggregated by year and across the eight primary categories covered by the platform. The results show a steady annual decrease in readability, suggesting that abstracts are likely becoming increasingly complex. Additionally, following the release of ChatGPT, a significant change in readability is observed for 2023 and the analyzed months of 2024. Similar trends are found across categories, with most experiencing a notable change in readability during 2023 and 2024. These findings offer insights into the broader changes in readability and point to the likely influence of AI on scientific writing.

*Keywords:* Readability; Scientific writing; LLMs; Generative AI;

# 1. Introduction

Since the release of ChatGPT in November 2022 and the subsequent emergence of tools enabling interaction with Large Language Models (LLMs), scientists have explored their impact on scientific writing and how generative AI tools, specifically LLMs, influence the way authors write, edit, and publish scientific papers (Kaebnick et al., 2023; Kendall & da Silva, 2024; Kousha & Thelwall, 2024; Lehr et al., 2024; Lund et al., 2023; Zielinski et al., 2023). Anecdotally, some scientists have shared that certain papers now have a distinct style, which they attribute to the use of these models by authors. Currently, there is no consensus within the scientific community on the appropriate use of these tools in this context. On one hand, some express concerns about the "soullessness" of text generated or edited by LLMs, as well as issues related to their training data and the potential for generating inaccurate information (Geng & Trotta, 2024; Lund et al., 2023; Walters & Wilder, 2023). On the other hand, some view these tools as a "great equalizer," enabling scholars with less proficient writing skills to enhance their productivity and meet the rigorous standards of high-quality journals and conferences (Berdejo-Espinola & Amano, 2023; Geng & Trotta, 2024). Additionally, several studies have investigated changes in scientific writing since November 2022, primarily focusing on shifts in word usage (Geng & Trotta, 2024; Kobak et al., 2024; Liang, Zhang, et al., 2024). This paper employs a different approach, readability metrics, to quantify changes in scientific writing following the release of ChatGPT.

Readability refers to how easily a reader can comprehend a text. Existing readability metrics use various techniques to quantify this aspect, often relying on formulas that assess elements such as the number of words, sentences, or syllables in a text. These metrics have been utilized to investigate topics related to scientific publications (Horbach et al., 2022; Sun et al., 2024; Vergoulis et al., 2019). For example, a study published in 2017 analyzed abstracts spanning several decades and used two readability metrics to quantify changes in readability, finding a consistent decline over time (Plavén-Sigray et al., 2017). Other studies have applied these metrics in different contexts, such as examining their correlation with citation counts (Ante, 2022; Dowling et al., 2018; McCannon, 2019) and assessing the readability of papers within specific disciplines (Dolnicar & Chapple, 2015; Graf-Vlachy, 2021; Wang et al., 2022). This paper builds upon recent research on the impact of generative AI, specifically LLMs, on academic research and publication, as well as longstanding scholarship that uses readability metrics to analyze changes in scientific writing. In this study, four readability metrics are used, each of which relies on features such as the number of characters, words, sentences, or syllables. Therefore, readability in this paper is determined based on these structural characteristics of abstracts. It is important to note that readability may be measured differently in other contexts or using alternative approaches.

The primary objective of this study is to use readability metrics to analyze changes in the readability of scientific abstracts over time, using a dataset of all papers posted to arXiv.org from 2010 to June 7, 2024. Specifically, this research is motivated by the hypothesis that the introduction and growing popularity of accessible generative AI tools, such as ChatGPT and Gemini, have contributed to a significant change in scientific readability. By assessing the readability of abstracts, this study aims to determine whether significant changes in readability have occurred following the release of ChatGPT, without exploring the causes behind these changes. A shift in measured readability could indicate that abstracts are becoming more complex, more difficult to read, longer, or more likely to include advanced terminology or jargon.

The main source of data for this work is arXiv.org. The platform enables scholars from eight broadly defined disciplines that include physics, computer science, and economics to host and share their work. These papers can be uploaded at various stages of the academic publishing process, and authors can note whether their papers have been published in a conference proceeding or journal. Information associated with abstracts such as their categories and upload dates help shape the research questions of this study. In summary, the study aims to aggregate readability scores across different variables, such as years and categories, to explore trends and changes in scientific writing.

More specifically, the objective is to address the following research questions:

- RQ1: What are the readability levels of all abstracts posted between 2010 and June 7, 2024, and are there any observed trends over this period?

- RQ2: Have there been significant changes in the readability of abstracts following the release of ChatGPT?
- RQ3: Are there differences in readability among the eight categories on arXiv, and how do these differences compare after the release of ChatGPT?

The rest of the paper is organized as follows: Section 2 includes a review of relevant related work. Section 3 describes the dataset and research methodology of the study. Section 4 contains the results of the study. Sections 5 and 6 present a discussion of the findings and a conclusion.

## 2. Related Work

Scientists have studied various aspects of how generative AI tools and LLMs have impacted or may influence scientific publishing in the future (Kendall & da Silva, 2024; Lehr et al., 2024). These efforts continue a long tradition of research aimed at understanding and improving the scientific community's engagement with rigorous scholarship (Abramo et al., 2020; Alsudais, 2021b, 2021a; Fraser et al., 2021; Wilkinson et al., 2016). Some studies have examined these models' effects beyond published papers, focusing on other aspects of publishing, such as using ChatGPT to assist in peer review (Hosseini & Horbach, 2023; Kousha & Thelwall, 2024). Similarly, one study investigated open academic peer reviews submitted to four top AI conferences, finding that "6.5% to 16.9% of text submitted as peer reviews to these conferences could have been substantially modified by LLMs, i.e., beyond spell-checking or minor writing updates" (Liang, Izzo, et al., 2024). Another related area of interest involves developing methods to detect papers written by these tools (Cingillioglu, 2023; Hamed & Wu, 2024).

Several studies have used the same arXiv.org dataset as this study to investigate research questions related to the use of ChatGPT and similar tools in scientific publishing. In one paper, the authors analyzed abstracts from this dataset and used various methods to quantify changes in word frequency since the release of ChatGPT (Geng & Trotta, 2024). They found that "ChatGPT is having an increasing impact on arXiv abstracts, especially in the field of computer science." They also identified specific words, such as "crucial," "significant," and "comprehensive," that have seen a substantial increase in usage since ChatGPT's release. In a related paper, the authors used the same dataset, along with others, and reported a "steady increase in LLM usage" (Liang, Zhang, et al., 2024). Their analysis highlighted specific terms with significant increases in usage, such as "pivotal," "showcasing," and "realm." Both studies indicated that computer science papers show a higher rate of LLM use compared to other categories, aligning with findings in this paper. Analyzing changes in word frequency has also been the focus of other studies, including research using PubMed abstracts published between 2010 and 2024 (Kobak et al., 2024). Additionally, some studies have focused on the impact of LLMs on publishing in specific disciplines such as dentistry (Uribe & Maldupa, 2024), veterinary neurology (Abani et al., 2023), and environmental psychology (Yuan et al., 2024). In the end, these studies reflect a growing interest in examining various aspects of generative AI, specifically LLMs, using abstract data and textual analysis. This paper contributes to this body of work by utilizing methods to assess changes in the readability of scientific texts following the release of ChatGPT.

## 3. Methodology

### 3.1 Dataset

The dataset used in this study is sourced from arXiv.org. According to a description on their website, arXiv is a "free distribution service and an open-access archive for nearly 2.4 million scholarly articles in the fields of physics, mathematics, computer science, quantitative biology, quantitative finance, statistics, electrical engineering and systems science, and economics" (arXiv, 2024). The platform is frequently used to host and share preprints before their official publication in conference proceedings or journals. Additionally, arXiv allows authors to update their submissions with journal references once their papers are published. This feature enables readers to distinguish between articles that remain unpublished in peer-reviewed outlets and those that have subsequently been published. Many researchers take advantage of this option to add journal or conference references to their articles. However, it's possible that some authors may not update their preprints with journal references after publication. Moreover, the accuracy of the journal references added to articles is not systematically verified. Some authors also upload their articles to arXiv after their papers have been accepted for publication, likely to increase visibility. In summary, papers available on arXiv represent a mix of unpublished preprints as well as preprints of peer-reviewed, published papers.

In 2020, Cornell University, which manages arXiv, announced that it would begin offering the entire dataset for download via the dataset and competition hosting platform, Kaggle (Cornell University, 2024). Since then, the dataset has been regularly updated and is available for download as a JSON file. This source is used to obtain the dataset for answering the research questions in this study. The dataset includes detailed information about each article, such as: the article ID, the submitter's name, the list of authors, the title, submitter comments, the journal reference, the DOI, the report number, the primary categories and subcategories (selected from a predefined list), the license, the list of article versions, and the last update date. For this study, the journal reference, category, abstract, and list of versions are used to address the research questions. Each article belongs to at least one of the eight primary categories mentioned earlier in this section, though papers may be classified under multiple categories. For example, an article can be categorized under both computer science and economics. Additionally, articles are assigned to subcategories. For instance, within the computer science category, subcategories include artificial intelligence and digital libraries, referred to as "cs.AI" and "cs.DL," respectively, in the dataset. Papers can also have multiple subcategories, meaning an article could be classified as both "cs.AI" and "cs.DL."

Since the dataset consists of papers posted to arXiv, an initial concern was determining the nature of these papers. Specifically, it was unclear whether the papers were primarily unpublished preprints or if they were versions of peer-reviewed, published papers. Previous research has attempted to explore this issue. In one study, the authors analyzed papers in the computer science category posted to arXiv between 2008 and 2017, discovering that "66% of all sampled preprints are published under unchanged titles, and 11% are published under different titles with other modifications" (Lin et al., 2020). However, the authors also noted that "the published preprints have significantly longer abstracts and introductions." Given that this paper aims to analyze the readability of abstracts from papers posted on arXiv and assess whether readability has changed since AI writing tools became more accessible, it is helpful to evaluate the similarity between these preprint abstracts and their corresponding published versions. This step helps determine whether the findings are limited to preprints.

One of the metadata fields in arXiv allows users to add journal references to their papers. Using this field, all papers with a reference to the *Journal of Informetrics* were retrieved. This journal was selected for this test mainly because it is the only journal containing "informetrics" in its title. A search using this term yielded 174 papers, all of which were confirmed to be published in the journal and not in any other publication with "informetrics" in the title. These papers were cross-referenced with the journal's official website, and the abstracts of the published versions were compared. The analysis revealed that 52 papers (32%) had identical abstracts in both versions. Additionally, many other papers had nearly identical abstracts, with only minor differences in punctuation, spacing, or a few words. Although this analysis is not the primary focus of the paper, it provides evidence that many papers on arXiv are not merely preprints but, in many cases, are identical to their published counterparts.

### 3.2 Processing the Dataset

The arXiv dataset was downloaded and processed to answer the research questions. However, several preprocessing steps were necessary to prepare the dataset and extract the required fields. One of the primary challenges was determining the publication dates of the papers, as the dataset does not have explicit columns for publication dates. Therefore, it was necessary to derive this information from other fields. Although the dataset includes an "updated date" column, this field seems to refer to the last time any information about the paper was modified, such as when an author adds a note or makes a minor revision. Therefore, other options to determine the dates for papers were considered. In arXiv, it is common for papers to have multiple versions, allowing authors to upload an initial version (v1) and later update it with newer versions (v2, v3, etc.). Each paper's list of versions is available in the dataset, and each version includes a full upload date, including the day, month, and year (e.g., [v1] Mon, 10 May 2021 09:32:21 GMT).

Every paper has at least one version, but some may have several. For this study, the latest version of each paper is used to determine the paper's publication date. The upload date of the most recent version is extracted and treated as the year of publication for the subsequent analysis. In all subsequent analyses, the publication year for each paper is determined based on the year of its latest version. To provide additional context on the potential effects and robustness of this design choice, all the papers in the dataset are also processed, and their readability results are quantified by assigning their publication dates to the date of the *first* version. In the results section, one aspect of the findings is presented using both approaches to assigning publication dates, and the readability results were found to be comparable and displaying similar trends.

The second major processing step involves extracting the categories for the papers. Each paper is classified into one or more of the eight super categories: computer science, economics, electrical engineering and systems science, mathematics, physics, quantitative biology, quantitative finance, and statistics. Since readability results are aggregated at the category level, identifying these categories is a required step. Seven of the categories follow a uniform format, making it straightforward to retrieve papers labeled under them. For instance, all computer science papers include "cs" followed by the ID of the subcategory, as described in the previous section. However, the physics category has multiple ID structures, requiring their formats to be compiled into a list. Papers labeled with any of these formats were then assigned to the physics category.

## 3.3 Readability Determination

With the extraction of dates and categories complete, the dataset is now ready for analysis. Several readability metrics are commonly used to assess the readability of texts. In this paper, four metrics are employed: 1) The Automated Readability Index (Kincaid et al., 1975), 2) The Coleman-Liau Index (Coleman & Liau, 1975), 3) The Flesch Reading-Ease (Flesch, 1948), and 4) the Flesch–Kincaid Grade Level (Kincaid et al., 1975). The formulas for calculating these metrics are listed in Table 1. The first test used in this study is the Automated Readability Index (ARI), which calculates readability based on the number of characters, words, and sentences in the text. ARI scores correspond to U.S. grade levels, with a score of 2 representing the first grade and a score of 13 representing the twelfth grade. However, scores can exceed these levels, reaching 15 or 16 for more complex texts. In this work, the objective is to track the evolution of ARI scores over time, regardless of their associated "grade level."

| Metric | Formula |
|---|---|
| ARI | $4.71 * \frac{\text{total number of characters}}{\text{total number of words}} + 0.5 * \frac{\text{total number of words}}{\text{total number of sentences}} - 21.43$ |
| CLI | $0.0588 * (\text{average of letters per 100 words}) - 0.296 * (\text{average of sentences per 100 words}) - 15.8$ |
| FRE | $206.835 - 1.015 * \frac{\text{total number of words}}{\text{total number of sentences}} - 85.6 * \frac{\text{total number of syllables}}{\text{total number of words}}$ |
| FKRGL | $0.39 * \frac{\text{total number of words}}{\text{total number of sentences}} + 11.8 * \frac{\text{total number of syllables}}{\text{total number of words}} - 15.59$ |

**Table 1.** The Four Readability Metrics and their Formulas.

The second test is the Coleman-Liau Index (CLI). Like ARI, CLI calculates readability by counting the number of letters, words, and sentences in a text, generating an estimate of the U.S. grade level required to understand the text. Specifically, CLI uses the average number of letters per 100 words and the average number of sentences per 100 words to produce its score. The third test is the Flesch Reading-Ease (FRE) test, which generates a score that reflects how easy or difficult a text is to read. A higher score indicates that the text is relatively easy to read, while a lower score suggests the text is more challenging. For example, a score above 90 suggests the text is extremely easy to read, whereas a score below 10 indicates a very difficult text. Although less common, extremely complex texts can even score below zero. As this test relies on counting syllables per word and words per sentence, it is possible that AI tools are generating texts with more complex words and longer sentences, leading to lower readability scores. The final test used in this paper is the Flesch–Kincaid Grade Level (FKRGL), which also calculates readability based on the number of syllables, words, and sentences. This test provides a score corresponding to a U.S. grade level, where a higher score indicates a more difficult text and a lower score suggests the text is easier to read, making it suitable for a lower grade level. One challenge when using FRE and FKRGL is that they are based on counting the number of syllables in words, which is more prone to errors compared to simply counting characters, words, and sentences.

In summary, the four tests used in this study to determine the readability of abstracts rely on counting characters, syllables, words, and sentences. The objective of using four separate tests is to ensure results that are robust and consistent. Additionally, the use of multiple tests is a common practice in similar studies employing comparable methodological approaches (Crossley et al., 2023; Plavén-Sigray et al., 2017). Three of the tests generate scores that estimate the U.S. grade level required to understand the text. For these three tests, an upward trend in average scores over the years would indicate that authors are writing more complex texts. In contrast, with FRE, an increasing complexity of texts is indicated by a downward trend in average scores, as the test assigns lower scores to more complex texts. It is important to reiterate that these tests rely on counting elements of text such as syllables, words, and sentences. Therefore, higher scores may simply reflect the presence of longer words or more complex sentence structures in the abstracts, which may not necessarily indicate that the abstracts are more difficult to read.

### *3.4 Answering the Research Questions*

Once the tests are selected, the next step is determining how to run them to produce readability scores for each paper. To calculate these scores, the Python library Textstat (Bansal & Aggarwal, 2024), which has been employed in other studies (Attia et al., 2023; Porwal & Devare, 2024), was used. The first research question (RQ1) focuses on investigating the readability scores and their evolution for papers uploaded between 2010 and 2024. The year 2010 was chosen as the starting point to provide a more focused analysis. The data for 2024 includes only papers from the first five months and the first few days of June (with a total of 6,904 papers posted in June 2024). To address the first research question, all abstracts were processed, and the values for ARI, CLI, FKRGL, and FRE were calculated using Textstat.

Each successfully processed paper was counted, allowing the averages for each test to be calculated at the end. Abstracts with fewer than 100 words were excluded, as the tests are recommended only for texts meeting this threshold. After processing, the averages for each test by year were calculated, resulting in four lists representing these yearly averages. Following this, Pearson's correlation coefficient was used to determine if statistically significant linear upward or downward trends exist. For each of the four lists, the test was applied, and p-values and Pearson's r were used to assess the existence of trends. Charts were then used to visualize and confirm that the relationship between a list and the years appeared linear. Finally, Shapiro-Wilk test for normality was used to determine if the variables used for each test were normally distributed. A high

Pearson's r (near +1) would indicate an upward trend for ARI, CLI, and FKRGL, while a low Pearson's r (near -1) would suggest a downward trend for FRE.

The second research question (RQ2) examines whether significant changes in readability occurred after the release of ChatGPT in November 2022. The objective was to identify any noticeable shifts, independent of their causes. This question was investigated using five distinct approaches. First, readability differences between each pair of consecutive years were calculated to identify any notable changes between 2022 and 2023, as well as between 2023 and 2024, compared to previous year pairs. Second, the percentage change between subsequent years was calculated. For example, if the average readability score was 10 in 2010 and 10.1 in 2011, the percentage change would be +1%. These percentage changes were then visualized to highlight any significant variations. The underlying hypothesis is that the largest percentage changes would occur in 2023 and 2024.

Third, the rolling standard deviation using three-year windows was calculated for all combinations of neighboring years. The purpose of specifying three-year windows was to identify the combination with the greatest increase in variability. The hypothesis is that the years 2022–2024 would exhibit the largest difference compared to all previous three-year combinations. Fourth, recent research that has already identified vocabulary changes in papers as a consequence of ChatGPT (Geng & Trotta, 2024; Kobak et al., 2024; Liang, Zhang, et al., 2024). One set of words identified as showing a significant increase in usage attributed to ChatGPT were "pivotal," "intricate," "realm," and "showcasing" (Geng & Trotta, 2024). The fourth approach to answering RQ2 involved analyzing all papers in the dataset published between 2020 and 2024 to identify whether any of the four specific words appeared in their abstracts. The same four readability metrics were then applied to compare abstracts containing any of these words with those that did not.

Finally, because papers can have multiple versions, those released in 2022 that had updated versions in 2023 or 2024 were identified. The objective of this approach was to measure changes in papers with multiple versions that were updated after the release of ChatGPT. Since ChatGPT was released on November 28, 2022, papers published in December 2022 were excluded. As the dataset only includes the abstracts from the latest versions of papers, it was necessary to retrieve earlier versions of the abstracts using alternative methods. Because many papers met this condition, it was not feasible to use the official arXiv API to retrieve abstracts for every older version as the rate limit policies for the API restrict retrieving the details of large number of papers[1]. Instead, three random samples were generated for testing. Three samples were generated, rather than a single sample, to obtain results that could be more reliably generalized to the entire set. To determine an appropriate sample size, a confidence level of 95% and a margin of error of 5% were used, resulting in samples of 380 papers.

For each paper in these samples, the arXiv API was used to retrieve the abstract of the older version, and the four readability measures were applied to calculate its readability scores. These scores were then averaged and compared with those of the updated versions. In cases where the older abstracts contained fewer than 100 words, the paper was excluded from the final calculations. Additionally, the same process was repeated for papers released in 2021 that had updated versions posted in 2022, excluding those updated after December 2022. The objective of this step was to compare the two sets of papers and determine whether those updated after 2022 showed a significant change in readability compared to those updated in 2022. A paired two-sample t-test was then used to determine whether a statistically significant difference exists between the values for each metric within each sample. In the end, all of these analyses to answering RQ2 were performed for each of the four metrics, with the goal of thoroughly investigating any major changes since November 2022. Ultimately, these five approaches were employed to determine whether significant changes in readability occurred in 2023 and 2024 compared to previous years.

The third research question (RQ3) compares the changes in readability across the eight categories since the release of ChatGPT. Due to data availability variations in some categories (e.g., two categories had only 11 papers between 2009 and 2011) and to maintain a focus on the period following November 2022, this analysis was limited to the years 2022, 2023, and 2024. For each category, the four test scores were calculated, and the percentage changes were determined in a manner similar to RQ2. A final analysis utilized paired two-sample t-tests to identify whether statistically significant changes occurred when comparing each set of metric values by category from one year to the next. For example, this included comparing the average ARI values for each category in 2022 to the corresponding ARI values for the same eight categories in 2023.

---

[1] https://info.arxiv.org/help/api/tou.html#rate-limits

## 4. Results

This results section describes the findings of this study. Each subsection is dedicated to answering one of the three research questions. Appendix A presents a detailed exploratory analysis of the dataset, focusing on the number of papers per year and the average length of abstracts.

### 4.1 Readability Results (RQ1)

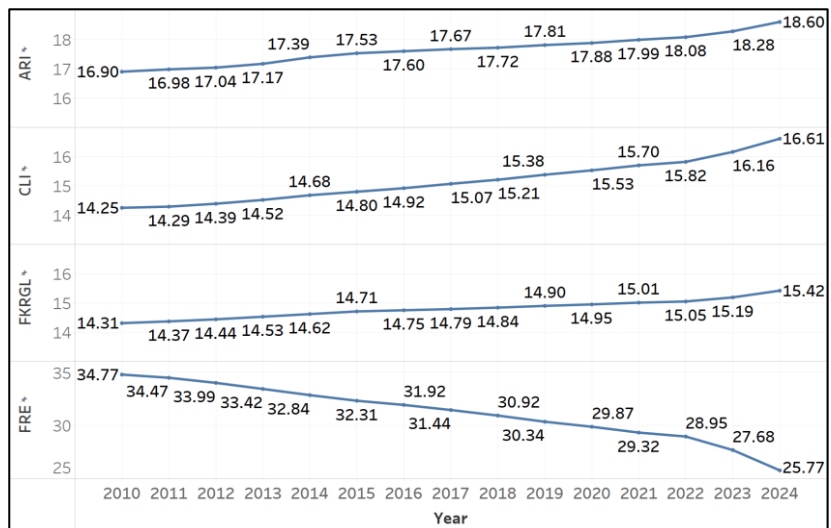

**Figure 1** Readability Results for Papers Per Year.

| Metric | Pearson's r | P-Value | S-W P-Value | Mean | Standard Deviation | Value for 2024 |
|---|---|---|---|---|---|---|
| ARI | 0.985 | <0.0001 | 0.914 | 17.6 | 0.491 | 18.6 |
| CLI | 0.982 | <0.0001 | 0.601 | 15.1 | 0.708 | 16.61 |
| FKRGL | 0.984 | <0.0001 | 0.984 | 14.7 | 0.31 | 15.42 |
| FRE | -0.982 | <0.0001 | 0.834 | 31.2 | 2.58 | 25.77 |

**Table 2.** Pearson Correlation Coefficient Results and Other Descriptive Statistics.

To investigate whether statistically significant trends exist, the Pearson correlation coefficient test was applied to each of the four metrics separately, using the yearly averages as input. Specifically, the input consisted of 15 entries (one average value per year from 2010 to 2024). For instance, to assess whether ARI exhibits a statistically significant trend, these 15 yearly averages were used as input. The results, along with descriptive statistics such as the mean and standard deviations, are presented in Table 2. The Pearson's r values were high and the p-values for all four metrics were below 0.001, indicating statistically significant upward trends for ARI, CLI, and FKRGL, and a statistically significant downward trend for FRE. Additionally, the results of the Shapiro-Wilk test indicated that the data for all four metrics were normally distributed, with all four showing high p-values (S-W P-Value in Table 2).

The results reveal several intriguing findings. First, across all metrics, scores increased each year, indicating that abstracts became more complex. It is important to note that for FRE, this increase is reflected by a decrease in scores, while for the other metrics, it corresponds to higher scores, as they are tied to grade levels. Second, the differences between the values in 2010 and 2024 suggest a major shift across all four metrics. Finally, while not immediately obvious in the chart, there appear to be significant changes between 2022 and 2023, as well as between 2023 and 2024. However, these changes are not definitive and are further explored in RQ2.

## 4.2 Investigating Changes Post ChatGPT's Release (RQ2)

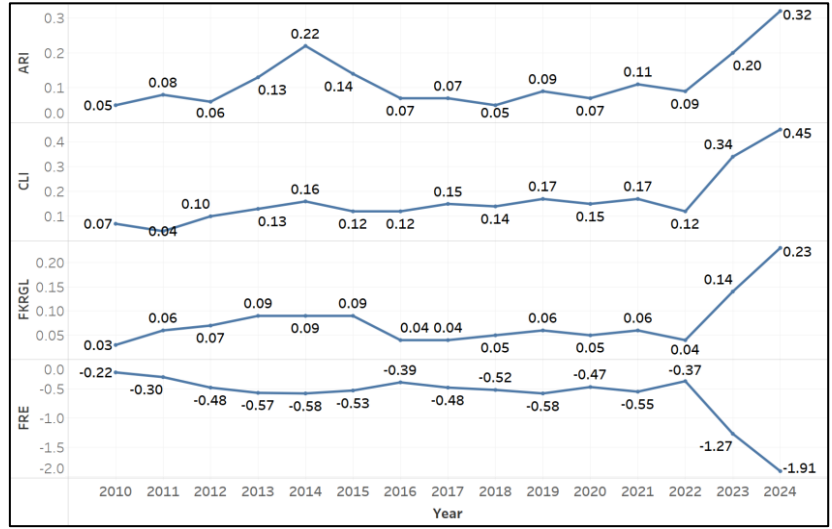

**Figure 2** Differences in Readability between Consecutive Years.

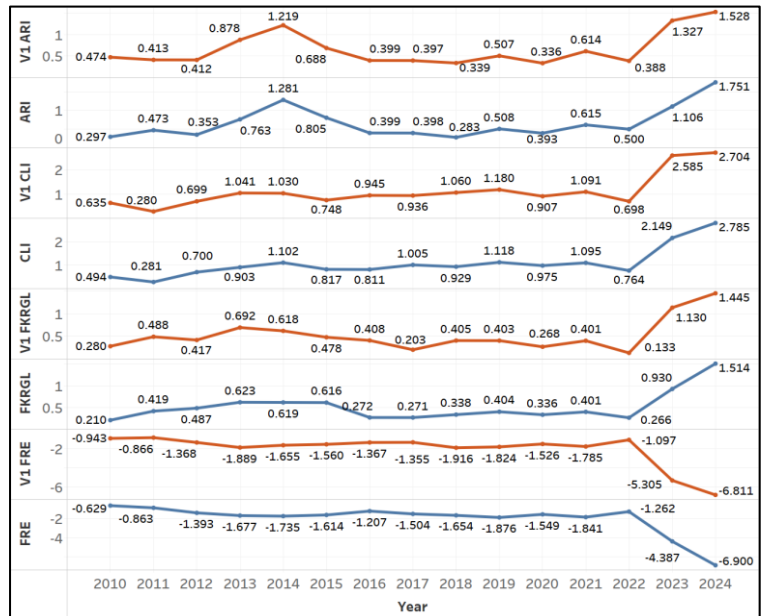

**Figure 3** Percentages of Change in Readability between Consecutive Years.

The second research question investigates whether significant changes occurred following the release of ChatGPT and the subsequent popularity of other accessible generative AI tools. To explore this, five approaches were examined: 1) differences in readability between pairs of consecutive years, 2) percentages of increase or decrease between pairs of consecutive years, 3) rolling standard deviation calculated using three-year windows as inputs, 4) differences between abstracts that include at least one word from a list of four words identified as increasingly used following the release of ChatGPT, and those that do not, and 5) changes in readability for papers with versions initially posted in 2022 and later updated in 2023 or 2024,

First, the differences in readability between each pair of consecutive years were calculated. The results are shown in Figure 2, with the values representing the differences between the current year and the previous year. For example, in 2022, the difference between the CLI value for that year and the previous year was 0.12. The results show a major change in readability in 2023 and 2024 across all metrics. A slight change in readability is observed between consecutive years before 2023, followed by a sudden significant change between 2022 and 2023, and again between 2023 and the processed months of 2024. To investigate this further, the percentages of change between consecutive years were calculated based on the raw readability values (not the differences). These results are presented in Figure 3. The findings were consistent, with the highest percentages of change occurring between the same two pairs of years. For example, the percentage of change in CLI between 2022 and 2023 was 2.149%. One notable exception is ARI between 2012 and 2014, where a sudden change in readability occurred, although the cause is unclear.

Figure 3 also presents the results when publication years were assigned to papers based on the date of their first version. Results for ARI, CLI, FKRGL, and FRE under this condition are labeled with "V1" and are displayed as orange line charts. For data without "V1" in the labels, the results correspond to publication years assigned based on the dates of the last version, consistent with the all the other findings discussed in this results section. The purpose of this comparison was to provide additional context and explore the effects of this methodological design choice of assigning publication dates for papers based on the dates of the latest versions. It is important to note that, while publication years for "V1" are assigned based on the first version's date, the abstracts analyzed are from the latest version of each paper. While the exact numbers differ, the trends remain consistent. Both methods show similar patterns: steady and uniform percentage increases across year pairs, followed by significant jumps from 2022 to 2023 and again from 2023 to 2024. In summary, the first two approaches to examining RQ2 demonstrated this pattern, providing evidence of significant changes occurring in 2023 and 2024.

The third analysis was performed by calculating the rolling standard deviation using three-year windows as inputs. For example, this was done for 2018, 2019, and 2020, and then again for 2021, 2022, and 2023. The differences in standard deviation can highlight the time periods that saw the largest increases in variability. For the four metrics, the two highest values were observed for the periods 2021–2023 and 2022–2024, with the only exception being the ARI values, where the second-highest was not 2021–2023, though the highest remained 2022–2024. This added further evidence on the significant changes in readability following November 2022.

The fourth approach focused on "pivotal," "intricate," "realm," and "showcasing," four words reported to have increased in usage following the release of ChatGPT (Geng & Trotta, 2024). To explore this, all papers in the dataset published in 2020, 2021, 2022, 2023, and 2024 were analyzed to determine the presence of any of these words in their abstracts. The results showed that 1.8% of papers in 2023 and 3.87% in 2024 included at least one of these four words, compared to just 0.63% in 2021 and 0.68% in 2022. To further explore the potential impacts of ChatGPT on scientific writing, the averages for the four readability metrics were calculated for each year, separately for papers containing any of the four words and for those without. In 2023, papers containing any of the four words had significantly higher values for the metrics, with an average ARI value of 19.4 compared to 18.25 for papers without these words. The overall average ARI for that year was 18.28, indicating that abstracts with any of the four words had a notably higher ARI value. Similar results were observed for the other metrics: CLI values were 18.0 for papers with the words compared to 16.1 for those without (overall average 16.6); FKRGL values were 16.16 compared to 15.16 (overall average 15.9); and FRE scores were 19.93 compared to 27.82 (overall average 27.68). While a similar trend was observed in 2021, the differences between the two groups were much smaller compared to 2023. For example, the ARI averages in 2021 were 17.9 for abstracts without the words and 18.6 for those with them. In 2024, the results mirrored those of 2023, with even larger differences between the two groups compared to 2021 and 2020.

These findings provide an additional layer of analysis and contextualize the results within the broader trends observed in previous research that focused on the potential impacts of ChatGPT and similar tools on scientific writing. A final approach to answering RQ2 examined papers with versions initially posted in 2022 that were later updated in 2023 or 2024. Since ChatGPT was released on November 28, 2022, papers with an initial version posted in December 2022 were excluded to avoid the possibility that their text could have been influenced by ChatGPT. This process yielded 23,836 papers. Three separate random samples were generated, and the arXiv API was used to retrieve the older abstracts. Then, all papers with a version in 2021 and an updated version in 2022 (excluding those updated in December 2022) were identified. The total number of papers meeting this condition was 22,122. Similarly, three random samples were generated, and their abstracts were compared. The objective was to 1) compare the average percentage changes for the four readability metrics between the two subsets and 2) use a paired two-sample t-test to determine whether a statistically significant difference exists between the values for each metric within each sample when comparing older abstracts to updated ones.

| Subset | Sample | Abstracts with a Change | Identical Abstracts | Older Abstracts <100 | ARI Difference | CLI Difference | FKRGL Difference | FRE Difference |
|---|---|---|---|---|---|---|---|---|
| Pre_2022 | 1 | 164 | 195 | 21 | 0.11% | 0.67% | -0.33% | -0.18% |
| | 2 | 177 | 181 | 22 | 0.72% | 0.97% | 0.61% | -1.9% |
| | 3 | 173 | 184 | 23 | 0.71% | 1.2% | 0.76% | -2.58% |
| Post_2022 | 1 | 188 | 166 | 15 | 2.23% | 1.61% | 2.62% | -7.04% |
| | 2 | 192 | 169 | 19 | 0.93% | 1.41% | 0.7% | -3.08% |
| | 3 | 195 | 172 | 13 | 0.39% | 1.09% | 0.59% | -3.23% |

**Table 3.** Samples used in Comparing Abstract Versions of the Same Papers and their Results.

| Subset | Sample | ARI Test Statistic | ARI P-Value | CLI Test Statistic | CLI P-Value | FKRGL Test Statistic | FKRGL P-Value | FRE Test Statistic | FRE P-Value |
|---|---|---|---|---|---|---|---|---|---|
| Pre_2022 | 1 | -0.15 | 0.87 | -1.30 | 0.19 | 0.44 | 0.65 | 0.09 | 0.92 |
| | 2 | -1.02 | 0.30 | -2.00 | 0.04 | -0.84 | 0.39 | 1.14 | 0.25 |
| | 3 | -1.1 | 0.27 | -2.4 | 0.01* | -1.15 | 0.25 | 1.64 | 0.10 |
| Post_2022 | 1 | -2.91 | 0.004** | -3.38 | 0.0009*** | -3.39 | 0.0008*** | 4.17 | 0.0000*** |
| | 2 | -1.25 | 0.21 | -2.57 | 0.01* | -0.91 | 0.36 | 1.77 | 0.07 |
| | 3 | -0.55 | 0.57 | -1.92 | 0.05 | -0.81 | 0.41 | 1.63 | 0.10 |

**Table 4.** T-tests Results Comparing Abstract Versions of the Same Papers. *p < 0.05; **p < 0.01; ***p < 0.001.

Table 3 presents descriptions of the samples and their sizes and differences between values for metrics for updated abstracts. In the table, the three samples listed under "Post_2022," indicates they were originally posted in 2022 and updated later, while "Pre_2022" refers to samples posted in 2021 and updated in 2022. The table also lists the number of papers with identical abstracts across the two versions examined. As shown in the table, many papers had abstracts that did not meet the length requirements for readability tests. For instance, in the first sample of papers initially posted in 2022 and updated in 2023 or 2024, 15 papers had initial abstracts that were too short to process. Additionally, the same sample included 166 papers with unchanged abstracts.

The table also displays the percentage differences in readability metrics between the older and updated versions of abstracts. For example, the values in the "ARI difference" column were calculated by first determining the ARI scores for both versions of the abstracts. The average ARI score for all older abstracts in a sample was calculated, and the percentage of change was then determined by comparing this average to the average score for the updated abstracts. For all four tests, the results indicated that the updated versions of abstracts had higher values compared to their older counterparts. For all metrics, the averages for "Post_2022" were higher than "Pre_2022." For example, for ARI, the average percentage change for the three samples of papers updated after 2022 was 1.18%, compared to only 0.51% for papers updated before 2022.

Following this, paired two-sample t-tests were performed to assess whether statistically significant changes occurred when comparing all abstracts in each sample to their updated versions. Table 4 presents these results, which show patterns similar to those observed in the previous analyses presented in Table 3. There was only one statistically significant result for the "Pre_2022" subset, while the "Post_2022" subset included several. However, the similarity between some samples from the different subsets, as well as the variability of results among samples within the same subsets, suggests that further analysis may be needed. Despite this, all four measures were higher on average for the subset of papers updated after 2022. In the end, these results and the four other approaches to exploring RQ2 provide evidence of significant changes in readability following the release of ChatGPT and the rise in popularity of similar generative AI tools. However, it is important to note that this is only one possible explanation, and other factors could have also contributed to this change.

## 4.3 Investigating the Categories (RQ3)

| Metric | Year | CS | ECON | EESS | Math | Physics | Q-Bio | Q-Fin | Stat |
|---|---|---|---|---|---|---|---|---|---|
| ARI | 2022 | 17.94 | 17.68 | 18.12 | 18.33 | 18.22 | 18.47 | 17.67 | 18.38 |
| | 2023 | 18.26 | 17.93 | 18.45 | 18.41 | 18.33 | 18.67 | 18.14 | 18.57 |
| | 2024 | **18.68** | **18.13** | **18.93** | **18.63** | **18.54** | **19.21** | **18.24** | **18.79** |
| CLI | 2022 | 16.35 | 15.68 | 16.62 | 14.94 | 15.55 | 16.68 | 15.51 | 16.28 |
| | 2023 | 16.78 | 16.12 | 17.01 | 15.12 | 15.75 | 17.08 | 16.08 | 16.53 |
| | 2024 | **17.33** | **16.38** | **17.54** | **15.39** | **16.01** | **17.64** | **16.39** | **16.83** |
| FKRGL | 2022 | 15 | 15.16 | 15.23 | 15.11 | 15.16 | 15.65 | 15.10 | 15.66 |
| | 2023 | 15.2 | 15.37 | 15.46 | 15.16 | 15.24 | 15.78 | 15.45 | 15.79 |
| | 2024 | **15.5** | **15.51** | **15.85** | **15.34** | **15.40** | **16.17** | **15.48** | **15.96** |
| FRE | 2022 | 27.79 | 28.23 | 26.08 | 30.60 | 29.42 | 24.50 | 29.19 | 25.09 |
| | 2023 | 26.11 | 26.29 | 24.49 | 29.98 | 28.63 | 23.11 | 26.70 | 24.00 |
| | 2024 | **23.75** | **25.22** | **21.77** | **28.80** | **27.45** | **20.53** | **25.91** | **22.69** |

**Table 5.** Readability Results the Eight Categories for the Years 2022, 2023, and 2024.

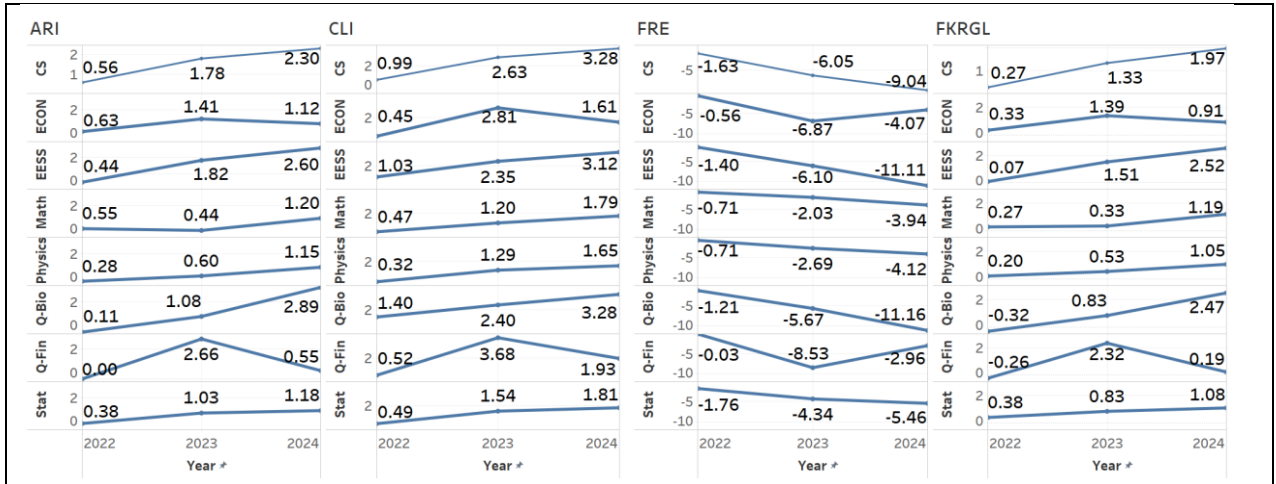

**Figure 4** Percentage of Increases in Readability Across the Eight Categories for the Years 2022, 2023, and 2024.

| Metric | Year | Test Statistic | P-Value |
|---|---|---|---|
| ARI | 2022 and 2023 | 5.40 | 0.001** |
| | 2023 and 2024 | 5.33 | 0.001** |
| CLI | 2022 and 2023 | 7.47 | 0.0001*** |
| | 2023 and 2024 | 7.71 | 0.0001*** |
| FKRGL | 2022 and 2023 | 5.09 | 0.001** |
| | 2023 and 2024 | 4.86 | 0.001** |
| FRE | 2022 and 2023 | -6.65 | 0.0002*** |
| | 2023 and 2024 | -6.05 | 0.0005*** |

**Table 6.** T-test Results Comparing the Eight Categories for the Years 2022, 2023, and 2024. *p < 0.05; **p < 0.01; ***p < 0.001.

The final research question focuses on investigating the individual eight super categories within the dataset. These categories vary significantly in popularity, with some showing few or no publications in the earlier years. The categories, ranked by the number of processed publications in 2023, are: Computer Science (CS), Physics, Mathematics (Math), Electrical Engineering and Systems Science (EESS), Statistics (Stat), Quantitative Biology (Q-Bio), Quantitative Finance (Q-Fin), and Economics (ECON). It is important to note that this analysis only reflects how these categories are represented in arXiv and may not fully represent the fields as a whole, since the platform's utilization varies across different research communities.

To answer this research question, papers from each category were retrieved, and the average readability metrics were calculated for each year and category. If a paper belonged to multiple categories (e.g., classified as both a physics and math paper), its readability metrics were included in the aggregated results for both categories. Upon reviewing the results, it became evident that several categories had few or no papers in the early years of the analysis. For instance, ECON and EESS together had only 11 papers processed between 2009 and 2011. This made it difficult to fairly compare the evolution of readability across categories over time. Therefore, to provide a clearer analysis, and to focus on changes post November 2022, only the results from 2022, 2023, and 2024 are presented here. The results are presented in Table 5. The results show that across the eight categories and the four readability metrics, scores increased each year, with the highest values for each category reached in 2024.

Following this, the analysis focused on the percentage change in readability for the categories, as illustrated in Figure 4. The values for 2022 represent the percentage change from 2021 to 2022. The results once again show greater changes for 2022-2023 and 2023-2024 compared to 2021-2022. These results, once again, demonstrate that most categories showed a significant shift in 2023 compared to the previous year, followed by another increase in 2024. The exceptions to this trend are the ECON and Q-Fin categories, where, although the highest values were achieved in 2024, the percentage change was greater in 2023 than in 2024. However, it is important to note that these two categories had the fewest papers with abstracts of at least 100 words. For comparison, in the processed months of 2024, there were only 1,343 ECON papers, 1,211 Q-Fin papers, and 65,102 CS papers.

Finally, each set of category scores for each metric in one year was compared to the corresponding set for the same metric in the following year (Table 6). Paired two-sample t-tests were performed twice for each metric, comparing 2022 to 2023, and 2023 to 2024, resulting in a total of eight tests. For example, for the ARI metric, the comparison between 2022 and 2023 involved testing the values for each category in 2022 (17.94, 17.68, 18.12, 18.33, 18.22, 18.47, 17.67, 18.38) against the corresponding values for 2023 (18.26, 17.93, 18.45, 18.41, 18.33, 18.67, 18.14, 18.57). The results indicate that all metrics showed statistically significant changes in values across categories. All eight comparisons yielded p-values below 0.01, with several falling well below 0.001. These findings provide further evidence of a significant shift in readability across most categories following 2022. Ultimately, the results for the three research questions highlight a substantial change in readability patterns after the release of ChatGPT.

## 5. Discussion

The results of this study demonstrate a steady change in the readability of scientific abstracts, as measured by established readability formulas, aligning with findings from previous work (Plavén-Sigray et al., 2017). Notably, this study also highlights significant shifts in readability in 2023 and 2024, as evidenced by the analyzed abstracts written after the release of ChatGPT. These findings contribute to the ongoing discussion surrounding the impact of generative AI, particularly LLMs, on scientific writing. This paper adds to this discussion by presenting evidence that the increasing prevalence of these tools is likely influencing the measurable readability of scientific abstracts. These changes may indicate that abstracts are becoming more complex or difficult. However, it is important to emphasize that this research does not seek to determine whether the rise of these tools directly caused an increase in the difficulty of scientific texts. Rather, it identifies statistically significant changes in readability metrics that emerged following the release of ChatGPT, regardless of the underlying causes. Other factors such as the use of advanced terminology or longer terms can produce higher readability scores. Future research that examines the specific nature of these changes could offer deeper insights into their scope and implications for scientific communication.

This research contributes to the ongoing debate on how scientific publishing should adapt to the growing presence and capabilities of these tools. It adds to the existing body of work examining the impact and ethical implications of using ChatGPT and similar technologies in academic writing, as scholars continue to explore and define appropriate use, responsible attribution, and broader ethical considerations. While LLMs can support some scholars in their writing, a concern is that not everyone has equal access to, or the ability to fully utilize, these tools. This raises important issues, as scholars who cannot afford or access LLMs may face added challenges in the publishing process. Therefore, investigating the accessibility of these resources and their impact on scholars' ability to publish warrants further attention. Still, this paper does not argue whether the changes in readability of scientific texts are inherently positive or negative. It is possible that this change is due to the accessibility of high-quality editing tools provided by AI, allowing scientists with less-than-ideal writing proficiency to improve the quality of the writing. On the other hand, more difficult texts may limit the accessibility of papers, as some readers may struggle with overly complex writing that is difficult to comprehend.

When considering all of this, it is important to note that LLMs respond to user prompts, and their behavior can be influenced by how they are directed (i.e. how these prompts are phrased). Although no prompts were used in this study, the way authors craft their own prompts to guide LLMs can have a direct impact on the edited texts generated by these tools. For example, when editing text, a prompt can specify that the user wants minimal edits and requests that the text not be made overly complicated or advanced. Therefore, perhaps one way to mitigate excessive increases in complexity is to recommend specific prompts for editing. As many journals now require the explicit declaration of AI use, these same outlets and publishers could explore recommended prompt lists to guide editing processes. Additionally, LLMs can be fine-tuned for specific tasks and datasets. Publishers and individual journals might consider developing their own fine-tuned models based on the type of writing they prefer. By training these models on the papers published in a specific journal, LLMs could learn to edit according to the typical difficulty and writing style historically featured in those journals. Journals can then provide these modified models to their potential authors. This approach could help manage the influence of AI on writing while creating valuable solutions that enhance both the quality and efficiency of the publishing process.

This paper has several limitations. First, the results emphasize the significant shift in readability since November 2022; however, only the first five months of 2024 and the first seven days of June are included in the analysis. It is unclear whether this partial data impacts the results. For instance, it is not known if abstracts posted in the first five months of the year typically have lower readability scores. In other words, once the rest of the year's data is processed, the 2024 averages may differ. Second, both FRE and FKRGL rely on counting syllables in words. In this study, Textstat was used to calculate all the readability metrics. Although this Python library has been used in other studies, it can make errors when calculating syllable counts. To address this, another library, Py-Readability-Metrics (DiMascio, 2020), was tested, but it also made mistakes in syllable counting. As a result, the findings for these two metrics may not be 100% accurate due to these issues. However, the primary objective is to track the evolution of the scores over time. Third, the data for each year may not always reflect the actual publication dates. For example, one author appeared to upload several papers in 2018 that were originally published between 2015 and 2018. Nevertheless, this issue is likely rare and insignificant given the consistency of the overall results. Fourth, this study focuses solely on the changes following the release of ChatGPT. Therefore, the findings may not generalize to other LLMs or to cases where ChatGPT is used together with other LLMs. Changes in readability may vary across different models, as some may alter text to different degrees. Additionally, examining the impact on readability when multiple LLMs are used is beyond the scope of the current study, and the release dates of other major LLMs were not considered. Finally, this study only uses one source of data. It is possible that the results are different once other sources are considered.

These limitations can still serve as starting points for future research. For instance, the entire analysis can be repeated after the end of 2024 to confirm whether the observed increase for 2024 remains consistent with the results presented in this study. Additionally, established scholarly datasets and resources can be leveraged to explore additional insights. For peer-reviewed papers, the spike in readability may appear more gradually and may not be immediately noticeable following November 2022 due to the slower pace of the peer review process compared to the immediacy of uploads to arXiv. Incorporating additional data sources could also strengthen the analysis by addressing confounding factors, which may improve the robustness of the findings and help reveal the actual impact of ChatGPT on the complexity of scientific texts. Similarly, considering the release dates of other major LLMs or analyzing their commonly used words could further enhance the analysis.

Other avenues for future research may consider countries and affiliations of authors and incorporate such information in the analysis. Similarly, a thorough examination of the changes observed across various disciplines that builds on the findings of this study could be beneficial. Such an investigation might reveal discipline-specific patterns of LLM usage and show that fields differ in how they leverage these tools. For example, a field like Computer Science, where authors may have adopted LLMs earlier than those in other disciplines, could exhibit early patterns of usage that are later replicated elsewhere. This could help trace the gradual adoption and utilization of LLMs within specific academic fields. Another potential avenue for future research is to explore alternative methods for measuring text complexity and readability, and to investigate whether their results align with those obtained using the four formulas applied in this study.

## 6. Conclusion

In this paper, a dataset of papers from arXiv posted between 2010 and June 7, 2024, was analyzed using four readability formulas to identify changes in readability over time, with a focus on determining whether significant shifts occurred following the release of ChatGPT. The results showed a steady increase in the complexity of abstracts, with notable shifts occurring after November 2022. Further analysis also highlighted the categories where these spikes were most prominent. In the end, this study contributes to the ongoing body of literature examining the potential impact of AI and LLMs specifically on how authors write and edit their work.

**Appendix A**

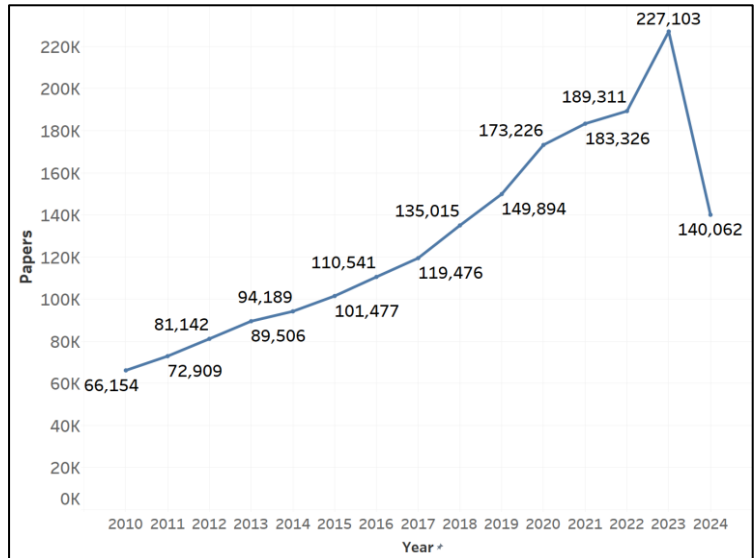

**Figure A.1.** Number of Papers Per Year.

Figure A.1 presents a line chart showing the number of papers in the dataset over time. As described in the methods, these numbers reflect the latest version of each paper, meaning the dates shown correspond to the latest version. For instance, if a paper was originally posted in 2010 but updated with a version 2 in 2011, it is counted in 2011. One initial observation from the results is the significant increase in the number of papers in 2023 compared to previous years. In 2021, the total number of papers was 183,326, which increased to 189,311 in 2022, a 3.26% rise. However, in 2023, the number of papers increased by 19.96% compared to 2022. This is the highest percentage increase between consecutive years, with the second largest being a 15.57% increase between 2019 and 2020. This notable jump in 2023 could be an early indicator of changes since the release of ChatGPT, though other factors may also have contributed or may be the primary reasons to this surge. In other words, multiple factors may have contributed to this change in numbers, and no single factor can be identified as the sole cause.

There were 140,062 papers dated 2024, but this figure includes only data up to the first seven days of June. Therefore, this number reflects papers from the first five months of the year, plus only the first seven days of June, during which 6,904 papers were posted. After excluding these June papers for the sole purpose of this estimation, the average number of papers per month in 2024 was 28,012. When multiplied by 12, the projected total for 2024 would be 336,148 papers, representing a significant increase compared to 2023. However, this estimation does not account for potential seasonal trends that could influence the average number of papers posted or updated each month. For instance, it is possible that the summer months see fewer papers on average compared to other times of the year. The 6,904 papers from June 2024 were included in subsequent analyses but were excluded only when estimating the total number of papers to be published in 2024.

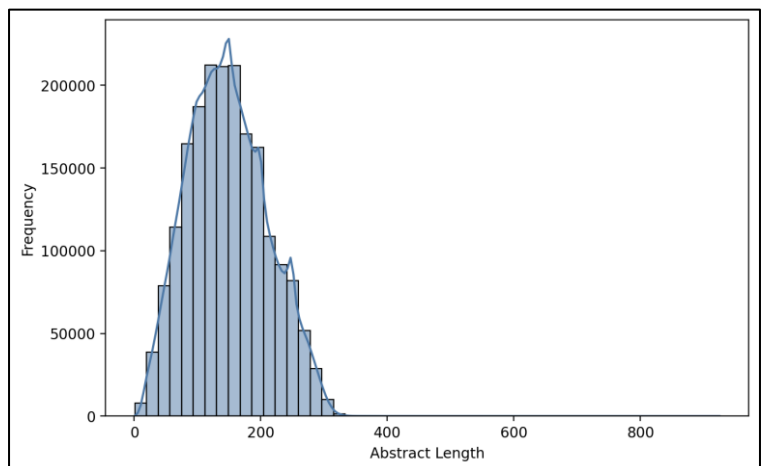

**Figure A.2.** Histogram of Abstracts' Length.

Following this, all abstracts were processed to generate initial statistics on their length. Figure A.2. presents a histogram of abstract lengths (in words) for all papers in the dataset. The number of bins was set to 50, and a kernel density estimate was applied to smooth the distribution. Papers with abstracts shorter than 100 words were excluded from subsequent analysis. During a review of these shorter abstracts, it was observed that some authors used the abstract space to provide brief updates on their papers. For instance, some indicated that they had discovered an error and were retracting the paper, or that they had replaced it with a new version, often including a link to the updated paper with a different title. The graph also reveals that few papers had abstracts longer than 400 words. Upon closer inspection, 142 papers had abstracts exceeding 400 words, 22 had abstracts longer than 500 words, and four papers had abstracts exceeding 600 words.